\documentclass[preprint]{aastex}

\tighten

\slugcomment{Accepted for publication in ApJ}

\newcommand {\kms} {$\,$km$\,$s$^{-1}$}
\newcommand {\cps} {$\,$ct$\,$s$^{-1}$}
\newcommand {\sol} {$_\odot$}
\newcommand {\nh}  {N$_{\rm H}$}
\newcommand {\sqig} {$\sim$}
\newcommand {\asca} {{\it ASCA}}
\newcommand {\rosat} {{\it ROSAT}}
\newcommand {\exosat} {{\it EXOSAT}}
\newcommand {\sax} {{\it BeppoSAX}}
\newcommand {\xte} {{\it RXTE}}
\newcommand {\axaf} {{\it Chandra}}
\newcommand {\src} {V382~Vel}

\shorttitle{X-ray emission from V382 Vel}
\shortauthors{Mukai \& Ishida}

\begin{document}

\title{The early X-ray emission from V382 Velorum (=Nova Vel 1999): \\
An internal shock model \\}

\author{Koji Mukai\altaffilmark{1}}
\affil{Code 662, NASA/Goddard Space Flight Center, Greenbelt, MD 20771, USA.}
\author{Manabu Ishida}
\affil{Institute of Space and Astronautical Science, 3-3-1 Yoshinodai,
Sagamihara, Kanagawa 229-8510, Japan.}

\altaffiltext{1}{Also Universities Space Research Association}

\begin{abstract}

We present the results of \asca\ and \xte\ observations of the
early X-ray emission from the classical nova V382~Velorum.
Its \asca\ spectrum was hard (kT$\sim$10 keV) with a strong
(10$^{23}$ cm$^{-2}$) intrinsic absorption.  In the subsequent
\xte\ data, the spectra became softer both due to a declining
temperature and a diminishing column.  We argue that this places
the X-ray emission interior to the outermost ejecta produced by
\src\ in 1999, and therefore must have been the result of a shock
internal to the nova ejecta.  The weakness of the Fe K$\alpha$ lines probably
indicates that the X-ray emitting plasmas are not in ionization equilibrium.

\end{abstract}

\keywords{stars: individual (V382 Vel) --- stars: novae, cataclysmic variables
--- X-rays: stars}

\section{Early X-ray Emission from Classical Novae}

Classical novae (or simply, novae) are explosions caused by
thermonuclear runaways on accreting white dwarfs (see, e.g.,
Chapter 5 of \citet{tome} for a review).  In common with many
other astrophysical explosions, a significant fraction of the energy
goes into the kinetic energy of the ejecta: for an ejecta
mass of 10$^{-4}$M\sol\ and an ejecta velocity of 1,000\kms,
one obtains \sqig 10$^{45}$ ergs as the ejecta kinetic energy;
these may be taken as typical values.
Of course, not all novae are identical; ``fast'' novae are visually
brighter at maximum, its visual light decays faster, and ejecta velocities
are higher, than the ``slow'' novae.  The fastness can be characterised
by the time it takes the nova to decline by 2 (t$_2$) or 3 (t$_3$)
visual magnitudes; there is a well-known correlation between the peak
absolute magnitude and the rate of decline, which makes novae useful
as distance indicators.  The white dwarf mass and other factors are known
to influence the fastness of a nova, although full details are still
being worked out.  Another important distinction can be discerned from
the abundances of the nova ejecta: roughly a third of recent novae are
neon novae, those believed to occur on O-Ne-Mg white dwarfs, while the
remainder are believed to occur on C-O white dwarfs.

The underlying binary is a cataclysmic variable (CV), that is,
a white dwarf accreting from a late type companion, usually
a Roche-lobe filling dwarf on or near the main sequence.  Under certain
conditions, the accreted material becomes degenerate; a sufficient
accumulation of this fresh fuel causes a thermonuclear runaway.
A nova typically reaches its peak visual brightness within a few days
after the onset of brightening.  In the early decay phase,
the intense wind from the still nuclear-burning white dwarf creates
a huge pseudo-photosphere, completely obscuring the underlying binary.
The declining mass-loss rate shrinks the photosphere, during which
the bolometric luminosity remains roughly constant, at about the
Eddington limit, and the effective temperature increases.  Finally,
when the photosphere has shrunk to the original radius of the white
dwarf, the nova may become a super-soft source, exhibiting an intense,
optically thick radiation from the white dwarf surface, with an
effective temperature of the order 50 eV.  Such super-soft emission
is observed 6 months to several years after the visual peak of the nova.
Recently, \citet{s2000} performed \axaf\ grating observations of
\src\ and V1494~Aql and discovered line-rich X-ray spectra, superimposed
on a super-soft continuum in the case of the latter but not the former.
These cast some doubt on the reliability of the parameters derived from
lower resolution X-ray observations (such as with \rosat\ PSPC), although
the gross characterization of the super-soft component is probably secure
in many cases.

In addition, an early, hard X-ray component has been observed in
several recent novae.  V838 Herculis (=Nova Herculis 1991,
V$_{\rm peak} \sim 5.0$) was detected 5 days past optical maximum
at 0.16\cps\ in \rosat\ PSPC \citep{ll92}.  V1974 Cygni
(=Nova Cygni 1992, V$_{\rm peak} \sim 4.2$) was detected 60 days past
maximum at 0.02\cps\ in \rosat\ PSPC \citep{b98}.  Nova Scorpii
1997 (V$_{\rm peak} \sim 9$) was detected \sqig 100 days past maximum
at 0.07 and 0.02\cps\ respectively in \sax\ LECS and MECS
\citep{o97}.  Finally, the \rosat\ PSPC detection
of V351 Puppis (=Nova Puppis 1991) 16 months after the visual maximum,
at 0.223\cps\ may also be due to the same component \citep{o96}.

Early X-ray emissions have also been detected from the recurrent nova,
RS~Ophiuchi, following its January 1985 outburst, using \exosat\ LE and ME
instruments \citep{ma86}.  The mass donor in this system is a red giant, unlike
in the short-period classical nova systems.  The early X-ray data
for this system have been interpreted in terms of the nova ejecta colliding
with the red giant wind.  In the classical nova systems with Roche-lobe
filling dwarf companions, any wind from the secondary would be too weak
for this mechanism to work.

In this paper, we report on the results of an \asca\ Target-of-opportunity
(TOO) observation and an \xte\ monitoring campaign of the early, hard X-ray
emission from \src.  Observations are described in \S 2, results are
presented in \S 3 and interpreted in \S 4.

\section{Observations}

\begin{figure}[th]
\begin{center}
\plotone{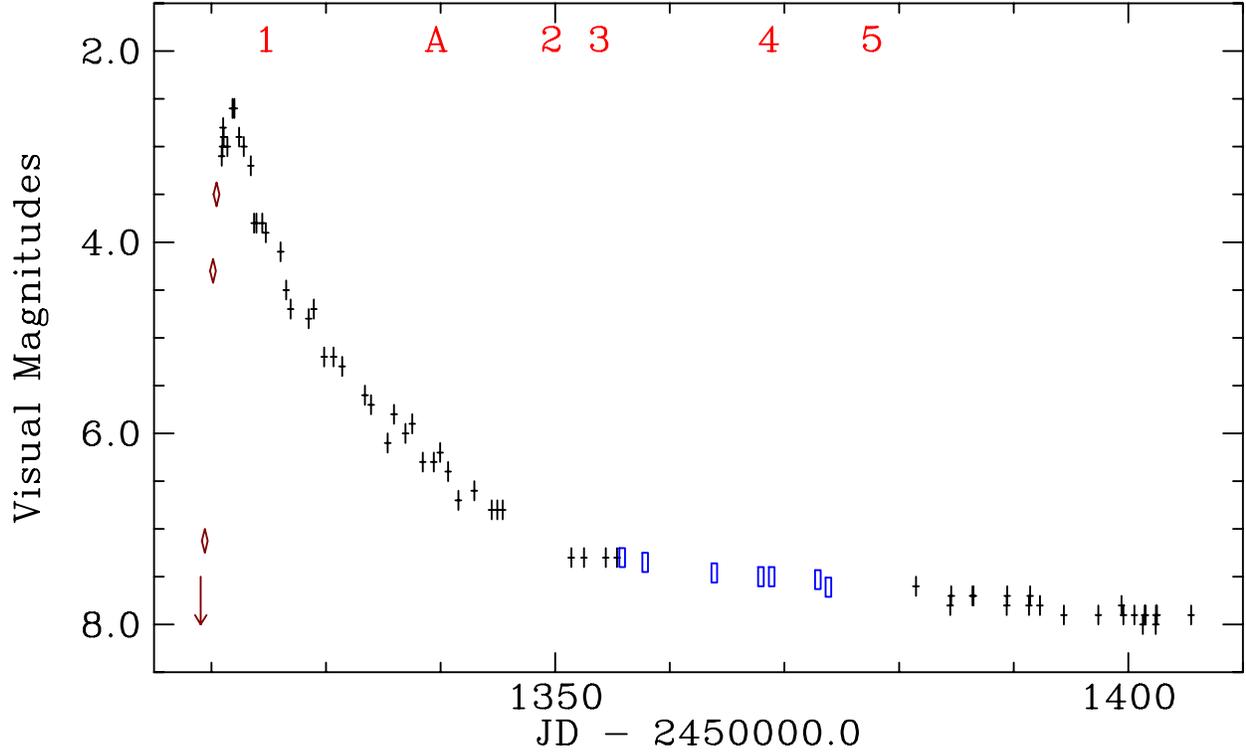}
\caption{The visual light curve of \src\ (crosses),
augmented by magnitudes estimated from prediscovery photographic
plates (an arrow indicating an upper limit of magnitude 13, and
diamonds indicating detections) and Mt. John University Observatory
photometry (open squares).  The times of X-ray observations that we
report in this paper are labeled with an ``A'' for the \asca\ observation
and by numbers for the \xte\ monitoring campaign.
}
\end{center}
\end{figure}

V382 Velorum (=Nova Velorum 1999) was discovered on 1999 May 22
at V \sqig 3.  The pre-discovery photographs extend
the detection back to May 20.923 UT at magnitude 7.0--7.25;
the nova was undetected at May 20.57 to a limiting magnitude
of \sqig 13.  We estimate that peak of thermonuclear runaway,
initial ejection of mass, and the beginning of visual brightening
all occurred around 1999 May 20.5 (=JD 2451319.0); we will refer to
this as time 0 of this nova in this paper.  Given the rapid rise,
this estimate is probably accurate to better than a day, which
is adequate for our purposes.

\src\ appeared to have peaked near $m_v \sim 2.8$
at about day 2.0, making this the brightest nova since V1500 Cygni
(=Nova Cygni 1975), and declined rapidly.  In Fig. 1, we have plotted
the visual magnitude estimates of \src\ as published in various IAU Circulars
(nos. 7176, 7177, 7179, 7184, 7193, 7203, 7209, 7236, and 7238),
since they provide the best overall coverage throughout the first
3 months of the nova.  We have supplemented these with magnitude estimates
from pre-discovery photographic plates, and with photometry at Mt. John
University observatory (\citet{k99} and \citet{g99}) between Jun 26 and
Jul 14 (a period for which no visual magnitude estimates are available
in IAUCs), although there could be an offset between visual magnitudes
and photographic or photoelectric measurements.  \src\ is a very fast nova:
\citet{d99} have measured the rate of decline of the nova to be t$_2$=6
days and t$_3$=10 days, and hence estimated a peak absolute visual magnitude
M$_{\rm V}$ of $-8.7 \pm 0.2$; this implies a distance to the nova of
about 2 kpc. It is also a neon nova \citep{w99}, as evidenced by the
detection of strong [NeII] 12.81$\mu$ line.

\begin{table}[th]
\begin{center}

Table 1. RXTE and ASCA Observations of V382 Velorum. \\

\begin{tabular}{llll}
\hline\hline
Date & Satellite$^a$ & Exposure (ksec)$^b$ & Count Rate $^c$ \\
\hline
1999 May 26 (day 5.7) & {\sl RXTE\/} (0123) & 2.4 & 0.11$\pm$0.03 \\
1999 Jun 9/10 (day 20.5) & {\sl ASCA\/} & 33.6/39.5 & 0.161$\pm$0.002/0.140$\pm$0.002 \\ 
1999 Jun 20 (day 31) & {\sl RXTE\/} (02)$^d$ & 1.3 & 3.59$\pm$0.06 \\
1999 Jun 24 (day 35) & {\sl RXTE\/} (0123) & 0.7 & 3.24$\pm$0.06 \\
1999 Jul 9 (day 50) & {\sl RXTE\/} (023) & 2.1 & 2.94$\pm$0.04 \\
1999 Jul 18 (day 59) & {\sl RXTE\/} (123) & 1.0 & 2.01$\pm$0.06 \\
\hline
\end{tabular}
\end{center}

$^a$For {\sl RXTE\/} observations, the PCUs that were
	used for the observations are indicated in parentheses.
$^b$Good on-source time after standard screening.
	For the {\sl ASCA\/} observation, exposures for GIS and SIS are shown.
$^c$For the {\sl RXTE\/} observations, 2.5--10 keV count rates
	per PCU are shown; for the {\sl ASCA\/} observation, the average
	GIS rate and the average SIS rate are shown.
$^d$This observation included a scan to confirm that the nova
	was the only source of hard X-rays.
\end{table}

The rare brightness of the nova (the brightest since the advent of
imaging X-ray astronomy) has made \src\ a prime target for X-ray
observations.  Accordingly, by the end of 1999, \src\ has been observed
with \xte\ (5 times), \sax\ (twice), \asca\ (once) and \axaf\ (once,
with three more pointings during 2000; \citet{s2000}).
Here we concentrate on the \xte\ and \asca\ data (summarized in Table
1; see also Fig. 1).  We also cite the preliminary results of the
\sax\ observations (\citet{o99a} and \citet{o99b}).

The \asca\ observation (see also the preliminary report by \citet{m99a})
was performed between 1999 June 9 13:09 UT and June 10 16:01 UT, for
approximately 40 ksec on-source.  We have performed standard data screening
and extraction, and combined the data from 2 pairs of similar instruments
for spectroscopic analysis (i.e., producing one SIS spectrum and one GIS
spectrum, each with an associated response and a background file).
For our light curve analysis,
we have combined the data from all 4 instruments.

There have been 5 public TOO observations of \src\ with \xte, from which we
have only analyzed the PCA data (a simple extrapolation of PCA spectral model
would argue against a HEXTE detection; even if a hard X-ray source was to
be detected, we cannot be confident of its true origin).  All were
performed with Epoch 4 gain setting, with varying number of Proportional
Counter Units (PCUs) on (see Table 1), obtaining usable data of \src\ for
0.7--2.4 ksec per visit.  In addition, during the middle of observation 2,
a raster scan was performed to confirm that \src\ is the only source of
hard X-rays in this area of the sky.  We have used the faint source model
for background subtraction (specifically, {\tt pca\_bkgd\_faintl7\_e04v03.mdl}
and {\tt pca\_bgd\_faint240\_e04e03.mdl} in addition to
{\tt pca\_saa\_history}).  We have used responses created by
{\tt pcarmf v7.01} for spectral fitting.

\section{Results}

\subsection{First \xte\ observation}

This observation was performed at day 5.7 in our convention, or only
about 3 days past the visual maximum.  Had a secure detection been
obtained, this would have been the earliest hard X-ray detection of
a classical nova.  However, this was not the case, as has been reported
earlier \citep{m99b}.  Even though there is a statistically
significant count excess over the background model in the 2.5--10 keV
band (our refined value is 0.11$\pm$0.03\cps\ per PCU), this cannot be
considered a secure detection, given the point-to-point fluctuation
in the cosmic X-ray background, particularly at such a low Galactic
latitude ($b^{II}= 5.8^\circ$).  A 0.2\cps\ per PCU source cannot be excluded,
roughly corresponding to 2.5$\times 10^{-12}$ ergs\,cm$^{-2}$s$^{-1}$
in the 2--10 keV band.

\subsection{\asca\ data}

The imaging capability of \asca\ leaves no doubt that \src\ was strongly
detected on day 20.5 (see also \citet{o99a} for the slightly earlier,
and equally secure, detection by \sax).

The combined 64-s bin light curve was analyzed for variability.
The best straight-line fit has a positive slope (combined count
rates increasing from 0.564\cps\ to 0.583\cps\ during our \sqig 1 day
observation), i.e., increasing with a timescale of \sqig 30 days
at day 20.5.  This fit has a $\chi^2_\nu$ of 1.15 for 626 degrees
of freedom, implying that the source was variable on a shorter
timescale formally at a 99.4\% confidence level.  However, a Fourier
analysis reveals no significant periodicity, to a limiting amplitude
of \sqig 5\%, and the apparent variability at this level may well be
due to imperfect background subtraction and other instrumental effects.

\begin{figure}[th]
\begin{center}
\plotone{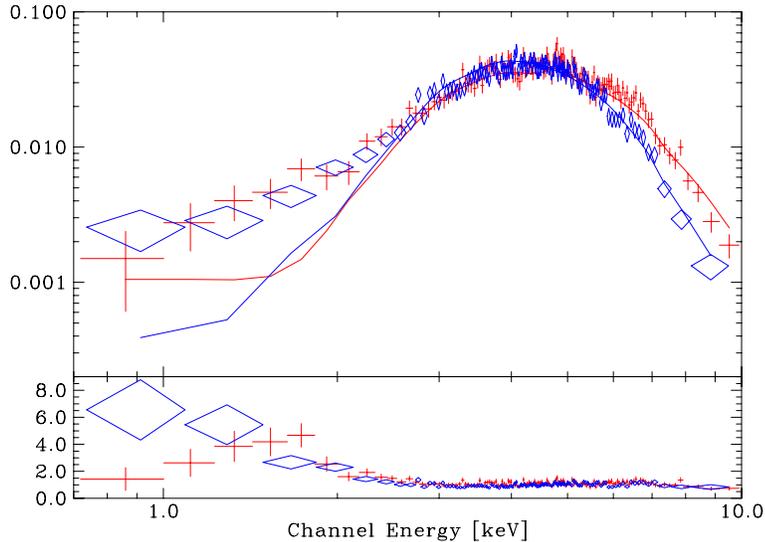}
\caption{The \asca\ spectra of \src.  GIS data (average of
GIS-2 and GIS-3) are plotted as crosses, and SIS data (SIS-0 and SIS-1
average) as diamonds, in the upper panel, also with the best-fit
bremsstrahlung model with a simple absorber.  In the bottom panel,
the residuals are shown in the form of data/model ratios.  A large
soft excess against this simple absorber model is clearly seen.  Also
apparent is the hardness of the intrinsic emission.  A weak Fe K line
is also visible in the upper panel.
}
\end{center}
\end{figure}

We have fitted the GIS and SIS spectra of \src\ with a Bremsstrahlung
continuum model (Fig. 2).  The choice of this model was dictated
by a combination of physical considerations and the quality of the fits,
not only of this observation but of the later \xte\ data as well.
We find that a single-component Bremsstrahlung model with a uniform
absorber gives a poor fit, with excesses at low energies.  A
partial-covering absorber model results in a marked improvement to
the fit: the absorbing column is found to be
1.01$\pm 0.05 \times 10^{23}$\,cm$^{-2}$, with a covering fraction of 99.5\%.
The bremsstrahlung model has a temperature of kT=10.2$^{+2.0}_{-1.7}$\,keV.
There is a weak detection a Fe K line at 6.63$\pm$0.11 keV with
an equivalent width of 130$^{+30}_{-70}$eV.  We have also attempted
fitting the \asca\ spectra with the {\tt mekal} plasma emission model
(\citet{m85}; \citet{m86}; \citet{l95}).
Since the continuum temperature is such that strong 6.7 and 6.97 keV
Fe K lines are expected, the fit fails unless the abundances are
allowed to vary; in this case, the abundance of Fe (the only element
the \asca\ data are sensitive to) of less than 10\% Solar is indicated.
The observed flux is 2.13$\times 10^{-11}$ ergs\,s$^{-1}$cm$^{-2}$
and inferred luminosity (corrected for absorption and for an assumed
distance of 2 kpc) is 4.5$\times 10^{34}$ ergs\,s$^{-1}$.  The inferred
emission measure (EM) is 1.7$\times 10^{57}$ cm$^{-3}$.

\subsection{Follow-up \xte\ observations: Spectral Evolution}

\begin{table}[th]
\begin{center}

Table 2. Results of Spectral Fits. \\

\begin{tabular}{lcccccc}
\hline\hline
Day & N$_{\rm H}$           & kT    & Line E & Line EqW & Luminosity     & EM \\
    & (10$^{22}$ cm$^{-2}$) & (keV) & (keV)  & (eV)     & (erg s$^{-1}$) & (cm$^{-3}$) \\
\hline
20.5 & 10.1$\pm$0.5 & 10.2$^{+2.0}_{-0.7}$ &
		6.63$\pm$0.11 & 130$^{+30}_{-70}$ &
		4.5$\times 10^{34}$ & 1.7$\times 10^{57}$ \\ 
31 & 7.7$\pm$2.0 & 4.0$^{+0.8}_{-0.6}$ &
		6.2$^{+0.2}_{-0.3}$ & 190$^{+120}_{-105}$ &
		7.9$\times 10^{34}$ & 6.3$\times 10^{57}$ \\
35 & 6.0$\pm$1.9 & 3.5$^{+0.7}_{-0.6}$ &
		6.4$\pm$0.4 & 220$\pm$150 &
		7.2$\times 10^{34}$ & 5.6$\times 10^{57}$ \\
50 & 3.1$\pm$1.2 & 2.5$\pm$0.3 &
		7.1$^{+0.6}_{-0.9}$ & 220$^{+230}_{-110}$ &
		7.6$\times 10^{34}$ & 5.9$\times 10^{57}$ \\
59 & 1.7$^{+2.4}_{-1.7}$ & 2.4$^{+0.6}_{-0.4}$ & & &
		4.9$\times 10^{34}$ & 4.0$\times 10^{57}$ \\
\hline
\end{tabular}
\end{center}
\end{table}

During the subsequent \xte\ campaign, \src\ was strongly detected,
and showed a marked softening from the \asca\ observation (day 20.5)
to the last \xte\ observation (day 59).  The spectra are shown in Fig. 3.

For these \xte\ spectra, we have used a simple absorber model, since
\xte\ PCA is not sensitive to the type of soft excess seen in the
\asca\ spectrum.  The column densities deduced from the fits
decrease to an almost undetectable (to \xte) level, accompanied
by a decrease in the temperature of the bremsstrahlung model;
either change in itself is not sufficient to explain the observed
spectral softening.  The Fe K line is securely detected above the
bremsstrahlung continuum model in observations 2 \& 3.  However,
even then, the lines are weaker than the plasma models would suggest.

These results are summarized in Table 2 and in Fig. 4.

\begin{figure}[th]
\begin{center}
\plotone{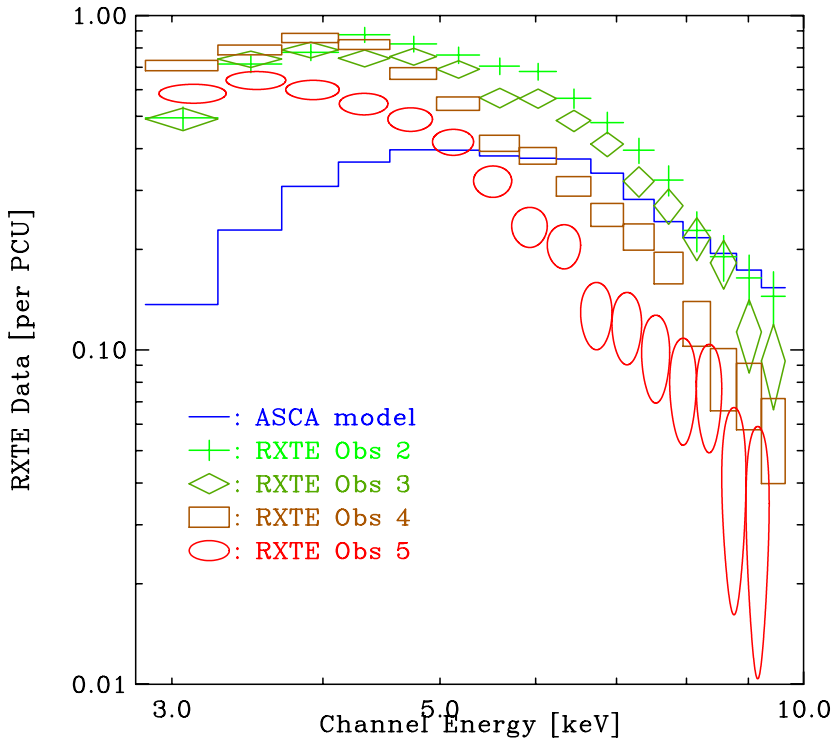}
\caption{The \xte\ spectra of \src.  The data from the
four observations in which \src\ was detected are plotted using different
symbols.  The histogram indicates the best-fit model from the \asca\ 
observation convolved with the \xte\ PCA response.
}
\end{center}
\end{figure}

\section{Discussion}

\begin{figure}[th]
\begin{center}
\plotone{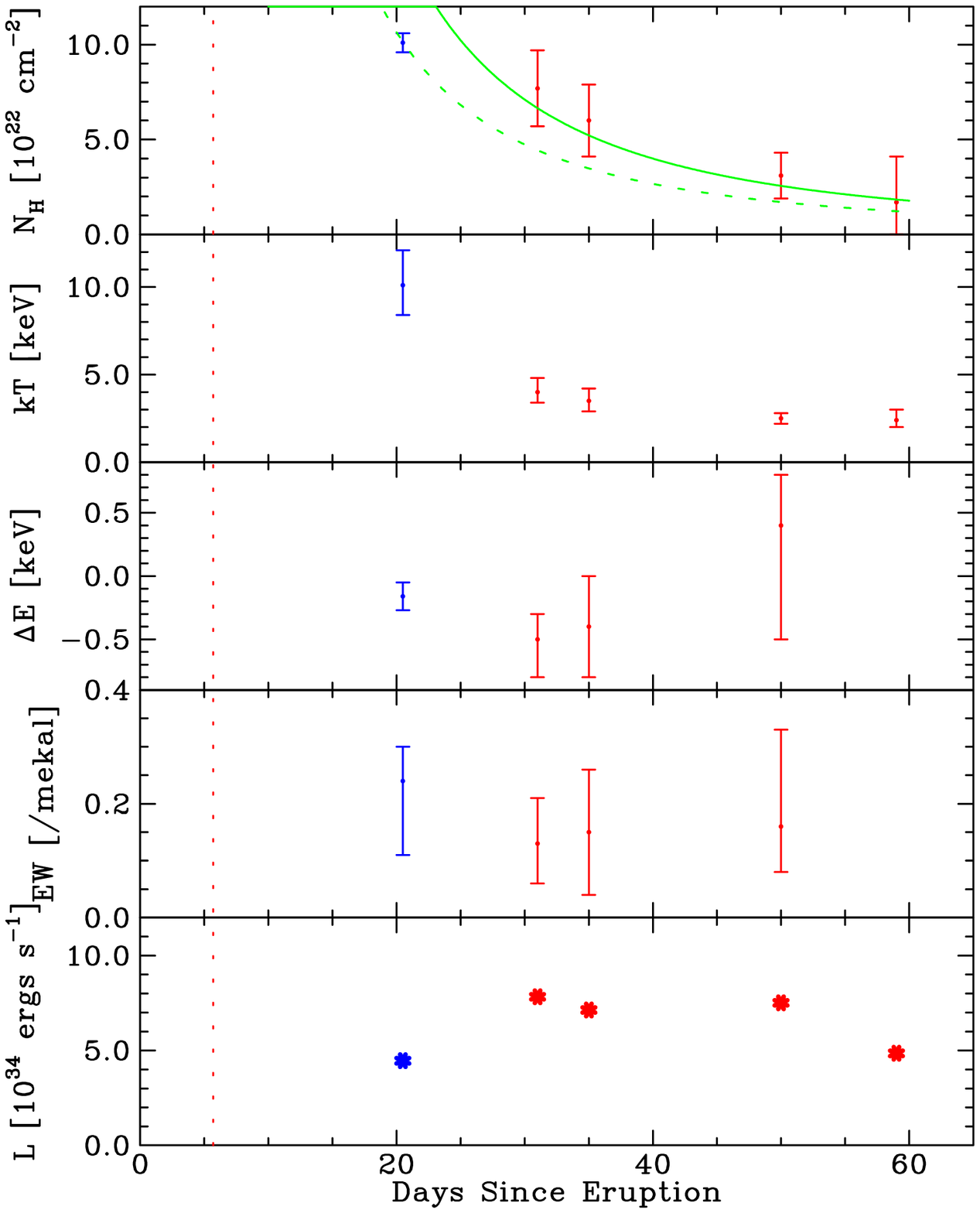}
\caption{The results of spectral fits of \src.
From top to bottom, the column density, the bremsstrahlung temperature,
the Fe K line energy shift relative to the plasma model prediction for
the appropriate continuum temperature, the observed-to-predicted
equivalent width ratio of the Fe K line, and the inferred total luminosity
for d=2kpc, are shown, as functions of time since eruption.  On the top
panel, the prediction of a simple $5.0 \times 10^{-5}$ (solid) and
$3.2 \times 10^{-5}$ (dotted) M\sol\ shell models are also shown.
}
\end{center}
\end{figure}

CVs long after a nova eruption are often seen as X-ray sources with
luminosities in the 10$^{30}$--10$^{34}$ ergs\,s$^{-1}$ range.  However,
it is unlikely that accretion can explain the X-rays we observed in \src,
given that the underlying binary was buried deep within the optically
thick wind at the epochs of these observations, quite apart from the
question of whether accretion could have been reestablished within
several weeks of the onset of the nova eruption.

Three nova-specific mechanisms for hard X-ray emission have been proposed:
radioactive decays, super-soft emission, and shock emission.  Radioactive
decays of $^{22}$Na produces 1.275 MeV $\gamma$-ray line, which could produce
X-rays via Compton-degradation \citep{li92}.  However, resulting X-rays
predominantly originate from a surface with Compton optical depth $\sim$1,
i.e., \nh\ $\sim 10^{24}$ cm$^{-2}$; this is far in excess of even the
highest column density seen in \src, $\sim$1.0$\times 10^{23}$ cm$^{-2}$
measured with \asca, hence we exclude radioactive decays from further
considerations.  The super-soft emission is the optically thick radiation
from the white dwarf surface, with an effective temperature of the order
50 eV, thus clearly of the wrong shape to explain our observations.
This component was, however, observed with \sax\ in November, 1999
in \src\ (\citet{o99b}).

This leaves shock emission as the only viable candidate as the origin
of X-rays from \src\ observed with \asca\ and \xte\ between day 20.5 and 59.

In one version of the shock model (see, e.g., \citet{ll92}), the nova
ejecta interact with pre-existing, circumstellar material.  There is
a severe problem with such an external shock model: as \citet{ll92}
point out, the presence of an unevolved secondary in most classical
nova makes the likelihood of extensive pre-outburst circumstellar
material low.  The amount of interstellar
medium that the nova ejecta can encounter during the first year is
obviously too small, and there is no evidence for a dense circum-binary
material in classical novae before an eruption.  As previously mentioned,
in a system with a red giant mass donor (such as RS~Oph), the external
shock model works well (\citet{ma86}; see also \citet{ll92}).  There
is no indication to date that \src\ has a giant mass donor, however.

We therefore discard the external shock explanation for the early
X-ray emissions in classical (non-recurrent) novae.

\subsection{Internal Shock Model}

The \asca\ and \xte\ observations reported here provide three valuable
clues as to the nature of the putative shock: the \nh\ history,
the kT history, and the behavior of the Fe K line.

The observed \nh\ cannot be interstellar, because it is variable.
Moreover, the UV observations \citep{sh99} indicate a reddening
E$_{\rm B-V}$ of perhaps 0.2, or \nh\ \sqig 3$\times10^{21}$ cm$^{-2}$.
This is also consistent with the November, 1999, \sax\ observation,
from which an X-ray column of \nh\ = 2$\times 10^{21}$ cm$^{-2}$
has been determined \citep{o99b}.

A simple model consisting of a discrete shell of mass 5$\times 10^{-5}$
M\sol\ expanding at 1000\kms, ejected at time 0 (the presumed
peak of the thermonuclear runaway), is successful in describing
the time history of \nh\ as measured with \xte\ if one assumes
a point-like X-ray source at its center.  For an extended source,
photons from near the limb has longer path lengths through the cold
outer shell, thus the above shell mass is an overestimate, particularly
for a limb-brightened X-ray source.  The point-source model that fits
the \xte\ observation overpredicts the \asca\ \nh\ by 50\%.  This may be
due, in part, to the simplistic treatment of the complex geometry; or
the outer shell may have been partially ionized at early stages, allowing low
energy photons to escape and hence complicating our spectral fits
(a similar mechanism may
have allowed the very early detection of V838~Her with \rosat\ \citep{ll92}).
Extrapolation of this model back to day 5.7, the epoch of the initial
\xte\ observation,  implies that the column would have been
$\sim 2 \times 10^{24}$ cm$^{-2}$, too high to allow X-rays escape even
allowing for some overprediction.  That is, the \nh\ history is suggestive
of an origin in an expanding shell, the ejecta from the 1999 nova eruption
itself.  The mass in this shell is probably somewhat less than
5$\times 10^{-5}$ M\sol\ if, as seems likely, the X-ray emission is
from a limb-brightened inner shell.

This model of the \nh\ history leads naturally to an internal shock model.
An expanding outer shell provides the observed \nh, with the X-ray producing
shock residing inside.  The simplest model, then, consists of two
distinct shells of nova ejecta.  The initial ejecta provide the absorbing
column; a layer of later, and faster-moving, ejecta plough into the initial
ejecta.  The high shock temperature of kT$\sim$10
keV requires a strong shock with velocity differential of $\sim$3000\kms.
Later observations show a softer spectrum, in kT as well
as in \nh, which suggests that the two sets of ejecta are merging to form
a single layer.  This is a scenario first proposed by \citet{f87},
which was motivated by the vast literature on the optical
spectra of classical novae in eruption.

Quantitative models of optical spectra of classical novae are generally
based on a single-component, optically thick wind approximation \citep{b76}.
However useful this formalism may be, it is clear from the rich
taxonomy of optical spectra of novae (summarized most notably by
\citet{pg57}) that nova ejecta are far more complex than this.
Several distinct systems are often recognized.  In time order,
these are called pre-maximum, principal, diffuse enhanced, and Orion
components.  As the name implies, the pre-maximum component is the first
absorption features seen, before the visual light curve reaches its maximum;
their typical velocities are in the 100--1000\kms\ range.
This component therefore is associated with the initial ejecta from
the nova eruption, which presumably carries the pseudo-photosphere with
it as it expands.  The principal system follows next, with a higher
velocity and a higher ionization; this is the system that persists
decades after the eruptions and can be identified with the expansion
velocity of the nova shells.  Diffuse enhanced and Orion systems are
yet of higher ionizations and higher velocities (1000\kms\ in slow nova
to as high as 4000\kms\ in very fast novae).

We see ever deeper into the optically thick wind as time goes by,
due to the decrease in the mass loss rate, hence the optical depth
at a given physical location.  This does explain the increasing
trend in ionization; however, we should observe less accelerated
material as time goes on in a one-zone wind model.  The fact that
the observed velocities
increase with time probably requires at least two distinct components.
For example, \citet{f87} explains the principal component as due
to the result of a collision between the slow-moving pre-maximum
system with the faster-moving diffuse enhanced/Orion system.
Applied to a fast nova, this model predicts a collision between
pre-maximum component moving typically at 1000\kms\ and the fast
wind with a typical velocity of 4000\kms, with the resulting shock
of kT$\sim$10 keV. This is just the X-ray temperature we observe
in \src.  Since we are unaware of any pre-maximum spectroscopy, we
simply adopt the ``typical'' value of 1000\kms; as for the fast wind,
\citep{sh99} report a terminal velocity of 5000\kms for Al III
$\lambda$1860 and Si IV $\lambda$1400 lines.  Thus the shock in
\src\ may have been capable of producing an even higher X-ray
temperature, but is consistent with the observed value.

In more theoretical terms, the pre-maximum system can be associated
with the dynamical ejection of the white dwarf envelop at near the
peak of the thermonuclear runaway; the faster materials can then be
associated with radiation-driven wind due to the continued shell
hydrogen burning, whose other manifestation is the super-soft
X-ray emission from the hot photosphere to be observed several month later.

Assuming that the shock is due to the collision between the pre-maximum
system and a fast-moving wind, what are the likely physical conditions?
First, the density of the pre-maximum ejecta can be estimated as follows.
Let us assume an ejecta mass of 2$\times 10^{-5}$ M\sol\ (as we have argued
that the 5$\times 10^{-5}$ M\sol\ figure from N$_{\rm H}$ history was likely
a slight overestimate).  The ejecta are expanding as a shell with radius
$v_{\rm pm} T$, where $v_{\rm pm}$ is the ejection velocity
\sqig 1000 km\,s$^{-1}$ and $T$ the time since explosion.  There is likely
to be a velocity dispersion $\Delta v$ (say, 200 km\,s$^{-1}$) in the
ejecta; taking the increasing radial spread into account, the volume of
the pre-maximum system at 20 days after eruption is
$\sim 1.6 \times 10^{43}$ cm$^3$ and the density is estimated to be
7.5 $\times 10^8$ cm$^{-3}$.

For the fast wind, since we assume this to be a continuous (and slowly
changing) phenomenon, velocity dispersions would not affect the density.
For a wind mass loss rate of 2$\times 10^{22}$ g\,s$^{-1}$ (or $\sim 10^{-4}$
M\sol\ in 100 days) at a wind velocity of 4000 km\,s$^{-1}$, the
wind density at the pre-maximum shell would be 7.5$\times 10^7$ cm$^{-3}$.
It appears likely that the fast wind will be initially assimilated into the
pre-maximum shell, after undergoing a shock, with a post-shock density
of the order 3$\times 10^8$ cm$^{-3}$.

If our interpretation is correct, the physical conditions of the
early X-ray emission source in \src\ is orders of magnitudes denser
than in supernova remnants, and orders of magnitudes more rarified
than in accretion shocks in CVs, two well-studied classes of shock
heated, X-ray emitting, plasmas.  Although it may be comparable
to stellar coronae in density alone, the heating mechanism and the
environment are different.  Applications of existing spectral models
(widely tested in supernova remnants and stellar coronae) must
therefore proceed with caution.

\subsection{Comparison with the Interacting Winds Model of \citet{o94}}

Such an internal shock model has already been suggested as a possibile
explanation of the early \rosat\ detection of V838~Her \citep{ll92}.
\citet{o94} has developed this into a detailed numerical model assuming a
constant mass loss rate, with ejection velocity of 1000 km\,s$^{-1}$
for the 1st day, increasing linearly to 3600 km\,s$^{-1}$ by day 5,
and remaining constant thereafter.  Our model and theirs are similar
in that X-rays are generated from an internal shock.  However, \citet{o94}
and we have chosen different sets of simplyfing assumptions.  \citet{o94}
assume a constant mass loss rate, with a smoothly changing ejection
velocity; in contrast, we have assumed a two distinct phases of mass loss
with a discontinuous change in velocity.  Are the differences significant,
and if so, which is the better framework on which to build future, more
detailed, models?

Let us first examine how the specific predictions of the \citet{o94}
numelical model compare with our data on \src: we find two significant
differences.

First difference concerns the predicted temperature of the X-ray emitting
region.  The \citet{o94} model predicts 2$\times 10^6$ to 2$\times 10^7$K
(or kT$\sim$0.2--2 keV) X-ray emitting plasma, matching one of the two
thermal plasma model parameters that fit the \rosat\ PSPC spectrum of
V838~Her (model RS2 in Table 2 of \citet{o94}), whereas we observe
KT$\sim$10 keV on Day 20.5 in \src.  Secondly, \citet{o94} claims
that ``for these parameter values the consequent reduction in low-energy
X-rays is small,'' whereas low-energy photons are decimated by intrinsic
absorption in the \asca\ spectrum of \src.

Note that, while the \citet{o94} model does predict a high temperature
region (T$\sim 10^8$K) 10 days after eruption, the density predicted in
this region is orders of magnitude too low to result in significant X-ray
emission (from their Figure 1, we estimate emission measure of order
10$^{46}$\,cm$^{-3}$, compared to $> 10^{57}$\,cm$^{-3}$ estimated from
\asca\ and \xte\ spectra).  Moreover, since this high temperature is seen
at the outermost edge of the ejecta, little absorption is predicted
(to be precise, half the emission should have interstellar \nh, while the
other half could suffer relatively high intrinsic absorption).

Thus, the specific numerical model presented in \citet{o94} cannot
explain our data on \src; however, this is not surprising.  We now ask
the more reasonable question: what changes in model parameters
might bring the predictions of the \citet{o94} model into line with
our observations of \src?

The penultimate paragraph of \S 3.2 of \citet{o94} makes it clear that
their model parameters are reasonably constrained by observations.
The mass loss rate is determined by the observed X-ray luminosity:
Given the similar X-ray luminosities inferred for V838~Her and \src,
we cannot siginificantly increase the mass loss rate adopted for the
former by \citet{o94}.  Without adjusting the mass loss rate, it does
not appear possible to match the \nh\ we observe in \src.  The temperature
of the X-ray emitting region in their model is determined by the velocity
contrast, $u_2-u_1$; because the line velocities measured in \src\ is
similar to those in V838~Her (\citet{v96} reports a terminal velocity
of 3000\kms\ and \citet{o94} assumed 3600\kms), we cannot arbitrarily
increase the velocity contrast, hence are unable to match the high
temperature (kT$\sim$10 keV) in \src.

It appears that there is not much room to adjust the parameters,
from their own analysis of their numerical model.  Given this,
we conclude that we probably need to abandon some of the simplifying
assumptions adopted by \citet{o94}.  We suggest that the simple
arguments we presented in \S 4.1 may serve as a starting framework
on which to construct detailed models, without the difficulties
encountered by the \citet{o94} version.

Finally, we note that any model that can succesfully explain \src\ may
apply, with minor modifications, to V838~Her.  This is because
the \rosat\ PSPC data on V838~Her can be fitted with a thermal model
with kT$>$4 keV (model RS1 in Table 1 of \citet{o94}).

\subsection{Weak Fe K Lines: Underabundant or Underionized?}

The weakness of the Fe K line is consistent with the shock model,
provided either that the Fe abundance is low in the ejecta, or that
the shocked plasma is not in ionization equilibrium.

There is no theoretical objection to a low Fe abundance in nova ejecta.
This could arise either because the white dwarf accretes low Fe abundance
material from the secondary, or because the heavy elements have settled down
to deeper layers of the primary in before the nova eruption.  However,
\src\ has been classified as a Fe II nova, because its optical
spectra include strong and broad Fe II lines (\citet{st99},
\citet{d99}).  In addition, the X-ray continuum shape in the 5--10 keV
region measured with \asca\ suggest the presence of an Fe edge at a level
consistent with a Solar composition absorber with N$_{\rm H} \sim 10^{23}$
cm$^{-2}$ (NB this is not a secure result on its own, as the Fe edge depth
is linked with the Fe emission line strength and the continuum shape in our
fit).  Clearly, Fe is present in the ejecta of \src.  Therefore, we prefer
to discount a low abundance as the explanation for the weak
Fe K features in the X-ray spectra.

Instead, we consider it likely that the Fe in the nova ejecta is
underionized.  Studies of supernova remnants typically find that
the ionization equilibrium is archived with a timescale $t$ such
that $nt \sim 10^{12}$ cm$^{-3}$s (e.g., \citet{m94}).  In young
supernova remnants with $nt < 10^{12}$ cm$^{-3}$s, the iron atoms
are in the process of being ionized and this can result in weaker
Fe K$\alpha$ line at an energy somewhat lower than at equilibrium.
We do not have a detailed model of how this might apply to \src,
and it may not be wise to apply the existing non-equilibrium models
of supernova remnants without carefully considering the different
conditions.

One observational constraint we have on the density is the
emission measure, which can be derived from the normalization
of the bremsstrahlung model: they are 1.7$\times 10^{57}$ cm$^{-3}$ for the
\asca\ data and $\sim 6 \times 10^{57}$ cm$^{-3}$ for the subsequent
\xte\ spectra.  If the emission region of volume $V$ has a uniform density
$n$, emission measure simply equals $n^2 V$.  We can calculate the minimum
density consistent with the observed X-ray spectrum by taking the volume
of the sphere within the 1000 km\,s$^{-1}$ pre-maximum ejecta front at
day 20 ($\sim 2 \times 10^{43}$ cm$^3$), and assuming a filling
factor of 1/4 (since a strong shock compresses by a factor of 4):
it is $\sim 2 \times 10^7$ cm$^{-3}$ (for the assumed 2 kpc distance).
Such a plasma will stay in nonequilibrium for \sqig half a day.  This
short timescale for reaching ionization equilibrium is a problem for
this interpretation: perhaps the observed X-ray emissions are dominated
by recently shocked materials.  On the positive side, there is a hint that
the detected lines were at lower energies than those predicted by the
ionization-equilibrium plasma models (Table 2 and Fig. 4), which is
predicted by the non-equilibrium models.  Clearly, we need higher quality
observations of future bright novae, as well as further modelling of nova
ejecta, to discover for certain the cause of weak Fe K$\alpha$ lines in \src.

\section{Conclusions}

We have observed early X-ray emission from a bright classical nova,
\src.  The X-ray spectrum was hard with kT$\sim$10 keV and
\nh\ = 1$\times 10^{23}$ cm$^{-2}$ 3 weeks after the onset of the
eruption, declining to 2.5 keV and $2 \times 10^{22}$ cm$^{-2}$
2 months after the peak in the optical.
Given an assumed distance of 2 kpc, \src\ maintained an X-ray luminosity of
$7.5 \times 10^{34}$ ergs\,s$^{-1}$ for at least 20, perhaps as long as
40, days.  The fluence in the X-ray component during this interval was
about $2 \times 10^{41}$ ergs, a small fraction of the estimated total
kinetic energy of the ejecta.

This evolving hard X-ray
emission can be best modelled as due to an internal shock within
the nova ejecta.  Such a shock was originally postulated by \citet{f87}
to explain the taxonomy of optical lines.  We have argued that a
detailed model developed by \citet{o94} to explain the \rosat\ PSPC
data on V838~Her cannot be adopted to explain our data on \src;
we have outlined our own model, broadly of the same type but with
different assumptions, that may serve as a starting point for future
modelling works.

Sensitive X-ray observations of other bright novae are necessary to clarify
the dependence of hard X-ray properties on the speed class.  However,
slower novae will almost certainly have a lower peak temperature
and remain obscured for a longer period.  Frequent optical spectroscopy
is also necessary to obtain the velocities of various ejecta components,
to be compared with the X-ray temperature evolution.  Perhaps most importantly,
the field of early X-ray emissions from classical novae is still in its
infancy such that a single bright and well-observed system can significantly
improve our level of knowledge, as we hope we have demonstrated in this
paper.  Given the presence of X-ray observatories of unprecedented
capabilities during the first decade of the 21st century, we can only
hope to see a dramatic increase in our knowledge.

\acknowledgments

We thank Prof. Nagase and others of the \asca\ operation team,
and Dr. Swank and others at the \xte\ project, for generous
allocation of target-of-opportunity times.  We also thank
the supernova remnant aficionados at the Laboratory of
High Energy Astrophysics for useful discussion concerining
the effects of non-equilbrium ionization.


\begin{thebibliography}{}

\bibitem[Balman et al (1998)]{b98} Balman, S., Krautter, J. \& \"Ogelman, H.
	1998, \apj, 499, 395.
\bibitem[Bath \& Shaviv (1976)]{b76} Bath, G.T. \& Shaviv, G. 1976,
	\mnras, 197, 305.
\bibitem[Della Valle et al (1999)]{d99}Della Valle, M., Pasquini, L,
	Williams, R. 1999, IAUC 7193.
\bibitem[Friedjung (1987)]{f87} Friedjung, M. 1987, \aap, 180, 155.
\bibitem[Gilmore (1999)]{g99} Gilmore, A.C. 1999, IAUC 7226.
\bibitem[Kilmartin (1999)]{k99} Kilmartin, P.M. 1999, IAUC 7216.
\bibitem[Liedahl et al (1995)]{l95} Liedahl, D.A., Osterheld, A.L.
	\& Goldstein, W.H. 1995, \apjl, 438, 115.
\bibitem[Livio et al (1992)]{li92} Livio, M., Mastichiadis, A., \"Ogelmann, H.
	\& Truran, J.W. 1992, \apj, 394, 217.
\bibitem[Lloyd et al (1992)]{ll92} Lloyd, H.M., O'Brien, T.J., Bode, M.F.,
	Predehl, P., Schmitt, J.H.M.M., Tr\"umper, J., Watson, M.G.
	\& Pounds, K.A. 1992, Nature, 356, 222.
\bibitem[Masai (1994)]{m94} Masai, K. 1994, \apj, 437, 770.
\bibitem[Mason et al (1986)]{ma86} Mason, K.O., C\'ordova, F.A., Bode, M.F.
	\& Barr, P. 1987, {\sl RS Ophiuchi (1985) and the Recurrent Nova
	Phoenomenon\/}, p167, ed. Bode, M.F., (Utrecht: VNU Science Press).
\bibitem[Mewe et al (1985)]{m85} Mewe, R., Gronenschild, E.H.B.M.
	\& van den Oord, G.H.J. 1985, \aaps, 62, 197.
\bibitem[Mewe et al (1986)]{m86} Mewe, R., Lemen, J.R.
	\& van den Oord, G.H.J. 1986, \aaps, 65, 511.
\bibitem[Mukai \& Ishida (1999)]{m99a} Mukai, K. \& Ishida, M. 1999, IAUC 7205.
\bibitem[Mukai \& Swank (1999)]{m99b} Mukai, K. \& Swank, J. 1999, IAUC 7206.
\bibitem[O'Brien et al (1994)]{o94} O'Brien, T.J., Lloyd, H.M. \& Bode, M.F.
	1994, \mnras, 271, 155.
\bibitem[Orio et al (1996)]{o96} Orio, M., Balman, S., Della Valle, M.,
	Gallagher, J. \& \"Ogelman, H. 1996, \apj, 466, 410.
\bibitem[Orio et al (1997)]{o97} Orio, M., Trussoni, E., Balman, S., \"Ogelman,
	H., Gallagher, J., de Martino, D, Della Valle, M., Gonzalez-Riestra, R.
	\& Salvelli, P. 1997, IAUC 6778.
\bibitem[Orio et al (1999a)]{o99a} Orio, M., Torroni, V., Ricci, R. 1999a,
	IAUC 7196.
\bibitem[Orio et al (1999b)]{o99b} Orio, M., Parmar, A.N., Capalbi, M.,
	Piro, L., Mineo, T. 1999b, IAUC 7325.
%\bibitem[Orio et al (2000)]{o00} Orio, M. et al 2000, submitted to \apjl.
\bibitem[Payne-Gaposchkin (1957)]{pg57} Payne-Gaposchkin, C.,
	{\sl The Galactic Novae\/}, (Amsterdam: North-Holland).
\bibitem[Shore et al (1999)]{sh99} Shore, S.N., Bond, H.E., Downes, R.,
	Starrfield, S., Gehrz, R.D., Krautter, J. \& Woodward, C.E. 1999,
	IAUC 7261.
\bibitem[Starrfield et al (2000)]{s2000} Starrfield, S., Shore, S.N.,
	Butt, Y., Drake, J., Bond, H.E., Downes, R., Krautter, J.,
	Wagner, R.M., Gehrz, R.D., Woodward, C.E., Della Valle, M.,
	Hauschildt, P.H. \& Truran, J.W. 2000, \baas, 32, 1253.
\bibitem[Steiner et al (1999)]{st99} Steiner, J.E., Campos, R. \& Cieslinski, D.
	1999, IAUC 7185.
\bibitem[Vanlandingham et al (1996)]{v96} Vanlandingham, K.M., Starrfield, S.,
	Wagner, R.M., Shore, S.N. \& Sonneborn, G. 1996, \mnras, 282, 563.
\bibitem[Warner (1995)]{tome} Warner 1995, {\sl Cataclysmic Variables\/}
	(Cambridge: Cambridge Univ. Press).
\bibitem[Woodward et al (1999)]{w99} Woodward, C.E., Wooden, D.H.,
	Pina, R.K. \& Fisher, R.S. 1999, IAUC 7220.

\end{thebibliography}
\end{document}